\documentclass[options]{JHEP3}

\usepackage{graphicx}
\usepackage{epsfig}

\title{Parikh-Wilczek Tunneling from Noncommutative
Higher Dimensional Black Holes}

\author{Kourosh Nozari$^{a,b}$ and S. Hamid Mehdipour$^{a,c}$\\

$^{a}$Department of Physics, Faculty of Basic Sciences,
University of Mazandaran,\\
P. O. Box 47416-95447, Babolsar, IRAN. \\

$^{b}$Research Institute for Astronomy and Astrophysics of
Maragha, \\P. O. Box 55134-441, Maragha, IRAN. \\

$^{c}$Islamic Azad University, Lahijan Branch,\\
P. O. Box 1616, Lahijan, IRAN. \\
E-mail: \email{knozari@umz.ac.ir}, \email{h.mehdipour@umz.ac.ir}}

\abstract{We study tunneling of massless and massive particles
through the smeared quantum horizon of the extra-dimensional
Schwarzschild black holes. The emission rate of the particles'
tunneling is modified by noncommutativity effects in a bulk
spacetime of dimension $d$. The issues of information loss and
possible correlations between emitted particles are discussed. We
show that even by considering both noncommutativity and braneworld
effects, there is no correlation between different modes of
evaporation at least at late-time and within approximations used in
the calculations. However, incorporation of quantum gravity effects
such as modification of the standard dispersion relation or
generalization of the Heisenberg uncertainty principle, leads to the
correlation between emitted particles. Although time-evolution of
these correlations is not trivial, a part of information coming out
of the black hole can be preserved in these correlations. On the
other hand, as a well-known result of spacetime noncommutativity, a
part of information may be preserved in a stable black hole
remnant.}


\begin{document}

\section{\label{sec:1}Introduction}

Radiation spectrum of an evaporating black hole which is almost
similar to black body radiation spectrum, can be described by a
characteristic temperature dubbed Hawking temperature. By adopting a
quantum field theoretical approach in curved spacetime, this
characteristic temperature is given by $T_H=\frac{\hbar
c^3\kappa}{2\pi k_BG}$, where $\kappa$ is the surface gravity that
demonstrates the strength of the gravitational field near the black
hole surface \cite{haw}. Unfortunately, Hawking's approach leads to
a non-unitarity of quantum theory which maps a pure state to a mixed
state as a result of purely thermal nature of the spectrum. A few
years ago, Parikh and Wilczek \cite{par1} suggested a new method
based on null-geodesics to extract the Hawking radiation via
tunneling across the event horizon. This method modifies the form of
the black hole radiation spectrum that is due to inclusion of
back-reaction effects. The tunneling process clarifies that the
modified radiation spectrum is not precisely thermal and this leads
to the unitarity of underlying quantum theory \cite{par2}.
Nevertheless, the form of the correction is not sufficient by itself
to recover information because of its failure to have correlations
between the tunneling probability of different modes in the black
hole radiation spectrum \cite{par3}. Recently, Nicolini, Smailagic
and Spallucci (NSS) \cite{nic}, in a new conceptual approach to
space noncommutativity, have shown that black hole evaporation
process should be terminated when black hole reaches to a minimal
mass, {\it i.e.} \texttt{black hole remnant} (for a review see
\cite{nic2}). This minimal mass is a consequence of existence of
minimal observable length \cite{mini1,mini2,mini3,mini4,mini5}. The
NSS viewpoint on coordinate noncommutativity is carried out by the
Gaussian distribution of coherent states and is consistent with
\texttt{Lorentz invariance}, \texttt{unitarity} and
\texttt{UV-finiteness} of quantum field theory \cite{sma}. Moreover,
noncommutativity of spacetime is an inherent property of the
manifold by itself even if there is no gravity. In this framework,
some kind of divergencies which arise in general relativity and
black hole physics, can be removed also. In a recent paper
\cite{noz1}, we have studied the Parikh-Wilczek tunneling through
the quantum horizon of a Schwarzschild black hole in noncommutative
(NC) spacetime. We have shown that, in the proposed NC setup there
is no correlation between different modes of the radiation which
reflects the fact that information does not come out continuously
during the evaporation process, at least at late-time. The
fundamental motivation of the present paper is to find an
alternative framework to recover the lost information in the black
hole evaporation process (see \cite{info} for comprehensive reviews
on existing approaches to resolution of {\it black hole information
paradox}). This alternative framework can be identified with the
correlations between different modes of evaporation. Another
question that we focus on is the possible impact of braneworld
scenarios and correlation between different emitted modes of black
hole evaporation. In other words, we want to know whether inclusion
of extra spatial dimensions in the framework of extra dimensional
scenarios can create correlation between different modes or not.
This is an important question which is considered here via the
tunneling framework of Parikh and Wilczek. With these motivations,
we apply back-reaction and noncommutativity effects to study black
hole evaporation process in a bulk spacetime of dimension $d$.
Recently, we have shown that incorporating quantum gravity effects
(such as Generalized Uncertainty Principle (GUP)
\cite{mini1,mini2,mini3,mini4,mini5} (see also \cite{gup,adl}) and
Modified Dispersion Relations (MDR) \cite{mdr1,mdr2,mdr3}) in the
black hole evaporation process has the capability to produce
correlations between different emitted modes \cite{hamid}. Although
the time evolution of these correlations are not obvious and deserve
further investigations, the possibility of realizing these
correlations itself is an important step toward the resolution of
black hole information paradox. Here we generalize this achievement
to our extra dimensional framework and discuss some related issues.

The paper is organized as follows: In Section \ref{sec:2} we
construct a noncommutative framework in the presence of extra
dimensions and we study evaporation of Schwarzschild black hole in
this setup. In Section \ref{sec:3}, a detailed calculation of
quantum tunneling near the smeared quantum horizon is provided. We
calculate tunneling probability of massless and massive particles
through the quantum horizon. The problem of lost information and
possible correlations between emitted particles are investigated.
The capability of the black hole evaporation process to produce the
correlations between different emitted modes via incorporation of
quantum gravity effects is investigated in Section \ref{sec:4}.
Summary and conclusions are presented in Section \ref{sec:5}.

\section{\label{sec:2}Noncommutativity in the Presence of Extra Dimensions}
Noncommutative quantum field theories (NCQFTs) have attracted a lot
of attentions during last two decades \cite{con} (see also
\cite{hin} for a purely phenomenological viewpoint on this issue).
There are also a lot of attempts to possible realization of these
effects on experimental grounds \cite{hew}. The NC spacetime
coordinates, in their simplest form, are deformations of usual
spacetime coordinates (with an arbitrary number, $d$ of spacetime
dimensions) in which operators $X^A$ representable by the Hermitian
operators $\hat{X}^A$, characterize a nonvanishing commutation
relation as follows
\begin{equation}
[\hat{X}^A,\hat{X}^B]=i\theta^{AB}=i\frac{C^{AB}}{\Lambda_{NC}^2},
\label{mat:1}
\end{equation}
where $\Lambda_{NC}$ is a characteristic energy or inverse length
scale related to the NC effects and $C^{AB}$ ($\theta^{AB}$) is a
dimensionless, real and anti-symmetric matrix, whose elements are
constant and are assumed to be of the order of one. The physical
explanation of the $\theta^{AB}$ is the smallest fundamental cell of
observable area in the $AB$-plane, in the same way as Planck
constant $\hbar$ explains the smallest fundamental cell of
observable phase space in quantum mechanics. The scale
$\Lambda_{NC}$, possibly and most reasonably is of the order of
Planck scale $M_{Pl}$. This is supported by the fact that most of
the phenomenological studies of the NC models have presumed that
$\Lambda_{NC}$ cannot lie far above the TeV scale \cite{hin,hew}.
Since the fundamental Planck scale in models with large extra
dimensions (LEDs) becomes as small as a TeV in order to solve the
hierarchy problem \cite{ant,ark}; it is possible to set the NC
effects in a TeV regime.

The framework of NCQFT is achieved by substitution of ordinary
products in commutative theory with \texttt{Moyal $\star$-products}
\cite{moy}. There exist many formulations of NCQFT based on the
Moyal $\star$-product that lead to failure in resolving of some
important problems, such as Lorentz invariance breaking, loss of
unitarity and UV divergences of quantum field theory. Unfortunately,
no flawless and completely convincing theory of noncommutativity
exists yet. Recently, the authors in Ref.~\cite{sma} explained a
fascinating model of noncommutativity, \texttt{the coordinate
coherent states approach}, that can be free from the problems
mentioned above. In this approach, General Relativity in its usual
commutative form as described by the Einstein-Hilbert action remains
applicable. If noncommutativity effects can be treated in a
perturbative manner, then this is defensible, at least to a good
approximation. Indeed, the authors in Ref.~\cite{cal} have shown
that the leading noncommutativity corrections to the form of the
Einstein-Hilbert action are at least second order in the
noncommutativity parameter $\theta$. The generalization of the QFT
by noncommutativity based on coordinate coherent state formalism is
also interestingly curing the short distance behavior of pointlike
structures \cite{nic,sma} (see also \cite{gru,riz}). In this method,
the particle mass $M$, instead of being quite localized at a point,
is described by a smeared structure throughout a region of linear
size $\sqrt{\theta}$. In other words, we shall smear the point mass
Dirac-delta function utilizing a Gaussian function of finite width.
So, instead of a delta function distribution, the mass density
distribution of a static, spherically symmetric, particle-like
gravitational source in $d$-dimensional bulk spacetime will be given
by a Gaussian distribution of minimal width $\sqrt{\theta}$ as
follows \footnote{Throughout the rest of this paper, natural units
are used so that $\hbar = c = k_B = 1$.}\cite{riz}
\begin{equation}
\rho_{\theta}(r)=\frac{M}{(4\pi\theta)^{\frac{d-1}{2}}}e^{-\frac{r^2}{4\theta}}.
\label{mat:2}
\end{equation}
The line element which solves Einstein's field equations $G_{BA} =
8\pi G_d T_{BA}$ in the presence of smeared mass sources can be
obtained as
\begin{equation}
ds^2=-\bigg(1-\frac{2{\cal{M}}_\theta}{r^{d-3}}\bigg)dt^2+
\bigg(1-\frac{2{\cal{M}}_\theta}{r^{d-3}}\bigg)^{-1}dr^2+r^2
d\Omega_{(d-2)}^2, \label{mat:3}
\end{equation}
with
\begin{equation}
{\cal{M}}_\theta=
\frac{m}{\Gamma(\frac{d-1}{2})}\,\int_0^{\frac{r^2}{4\theta}}dt\,e^{-t}
\,t^{(\frac{d-3}{2})}\,\,,\,\, m = \frac{8\pi G_d M}{(d-2)
\Omega_{(d-2)}}. \label{mat:4}
\end{equation}
Where $d\Omega^2_{(d-2)}$ is the line element on the
$(d-2)$-dimensional unit sphere and $d$ is spacetime dimensionality.
The volume of the $(d-2)$-dimensional unit sphere is given by
\begin{equation} \Omega_{(d-2)} =
\frac{2\pi^{\frac{d-1}{2}}}{\Gamma(\frac{d-1}{2})}.\label{mat:5}
\end{equation}
$G_d$ is gravitational constant in $d$-dimensional spacetime which
in ADD \cite{ark} model is given by
\begin{equation}
G_{d} = \frac{(2\pi)^{d-4}}{\Omega_{d-2}}M_{Pl}^{2-d},\label{mat:6}
\end{equation}
where $M_{Pl}$ is the $d$-dimensional Planck mass and there is an
effective 4-dimensional Newton constant related to $M_{Pl}$ by
\begin{equation}
M_{Pl}^{2-d}=4\pi G_{4}R^{d-4},\label{mat:7}
\end{equation}
where $R$ \,is the common radius of all $(d-4)$-toroidally
compactified extra dimensions. We note that in this work, the
conventions for definition of the fundamental Planck scale $M_{Pl}$
are the same as which have been used by ADD (and also by authors of
Ref. \cite{gt}). The metric (\ref{mat:3}) smoothly interpolates
between de Sitter core around the origin and an ordinary
Schwarzschild geometry far away from the origin \cite{nic,riz}. On
the other hand, the curvature singularity at the point $r=0$ is
eliminated by noncommutativity as an inherent property of the
manifold. In this situation, a regular de Sitter vacuum state will
be formed accounting for the effect of the NC coordinates
fluctuations at short distances. Also, this scenario tends to usual
Schwarzschild spacetime at large distances. In the limit
$\theta\rightarrow0$, one recovers the complete Gamma function,
$\int_0^{\frac{r^2}{4\theta}}dt\,e^{-t}
\,t^{(\frac{d-3}{2})}\rightarrow\Gamma(\frac{d-1}{2})$, and
${\cal{M}}_\theta\rightarrow m$ as expected. Taking a 4-dimensional
viewpoint in the commutative case, one can simply obtain
${\cal{M}}_\theta=G_4M$. Then the NC (modified) Schwarzschild
solution reduces to the commutative (ordinary) case. Depending on
the different values of $M$, $d$ and $M_{Pl}$ and within a numerical
procedure, the metric (\ref{mat:3}) displays three possible causal
structure \cite{nic,noz1,riz,noz3} (see also \cite{nic3,nic4}):
\textbf{1}- It is possible to have two distinct horizons
(Non-extremal Black Hole), \textbf{2}- It is possible to have one
degenerate horizon (Extremal Black Hole), and finally \textbf{3}- It
is impossible to have horizon at all.

\begin{figure}[htp]
\begin{center}
\includegraphics{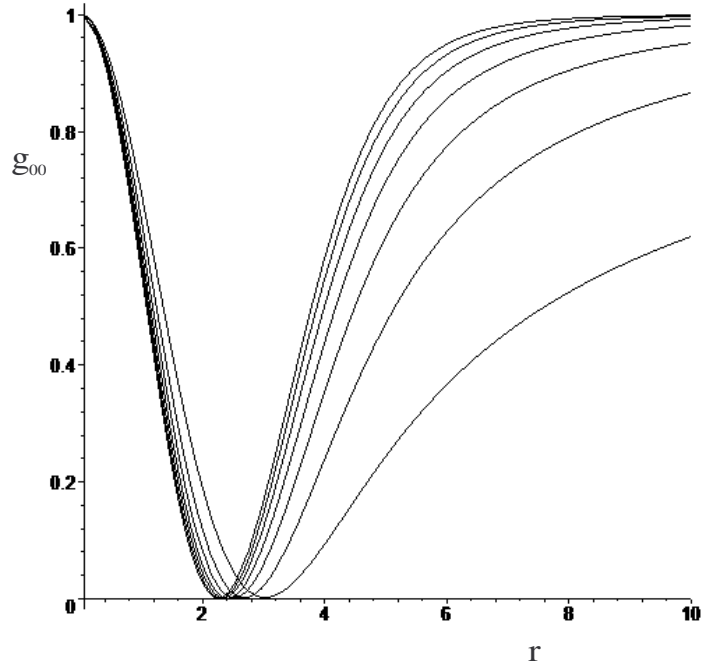}
\end{center}
\vspace{7.1 cm} \caption{\scriptsize {$g_{00}$ versus the radius $r$
for different values of black hole mass $M$ and different number of
spacetime dimensions, $d$. The figure shows the possibility of
having extremal configuration with one degenerate event horizon
(Extremal Black Hole) at $M=M_0$. This shows the existence of a
minimal non-zero mass that black hole shrinks to. On the right-hand
side of the figure, curves are marked from bottom to top by $d = 4$
to $d=10$ (with $M_{Pl}=1$). }} \label{fig:1}
\end{figure}

The radiating behavior of such a modified Schwarzschild black hole
can now be easily investigated by plotting the temporal component of
the metric, $g_{00}$, versus the radius $r$ for an \texttt{extremal
black hole} with different values of $M$ and $d$. This has been
shown in Fig.~(\ref{fig:1}).
\begin{table}
\caption{Minimal non-zero mass of the black hole (remnant mass) for
different number of spacetime dimensions.}
\begin{center}
\begin{tabular}{|c||c|}
\hline
\multicolumn{2}{|c|}{Extremal Black Hole\,\,;\,\,$M_{Pl}=1$\,TeV} \\
\hline $d$ & Minimal Mass (TeV) \\
\hline$4$ & $M=M_0\approx1.9$ \\
\hline$5$ & $M=M_0\approx3.94$ \\
\hline$6$ & $M=M_0\approx5.43$ \\
\hline$7$ & $M=M_0\approx5.77$\\
\hline$8$ & $M=M_0\approx5.08$\\
\hline$9$ & $M=M_0\approx3.88$\\
\hline$10$ & $M=M_0\approx2.65$\\
\hline$11$ & $M=M_0\approx1.64$\\
\hline
\end{tabular}
\end{center}
\label{tab:1}
\end{table}
As this figure shows, the coordinate noncommutativity leads to the
existence of a minimal non-zero mass in which black hole (due to
Hawking radiation) can shrink to {\footnote{For simplicity of
numerical calculations, we assume $\theta=1$. Naturally
$\frac{1}{\sqrt{\theta}}$ ($\Lambda_{NC}$) points out the energy
threshold beyond which a particle moves into the deformed spacetime
in which quantum gravity effects become important.}. It should be
emphasized that based on our computations, if we set $d=11$, we find
$M=M_0\approx1.64$ TeV, {\it i.e.} the minimum value of black hole
mass reduces to a value less than its value for $d=4$ (see
Table~\ref{tab:1}). Therefore, in theories with large extra
dimensions, if the number of spacetime dimensions becomes
sufficiently great, {\it e.g.} $d\geq11$ with a sufficiently small
NC parameter ($\Lambda_{NC}\sim M_{Pl}\sim 1$ TeV), then the
noncommutativity effect can enhance the possible formation and
detection of black holes in TeV-scale collisions at the LHC
{\footnote{ A micro-black hole in theories with large extra
dimensions can be produced at the large hadron collider (LHC) just
in the situation that $E_{cm}>M_0$, where $E_{cm}$ is {\it
parton-parton} center-of-mass energy which is equal to $14$ TeV.}}
\cite{arg}.

The event horizon radius,\, $r_H$,\, can be obtained from the
equation $g_{00}\left(\, r_H\,\right)=0$, which gives
\begin{equation}
1 - \frac{2{\cal{M}}_\theta(r_H)}{r_H^{d-3}}=0,\label{mat:8}
\end{equation}
with
$${\cal{M}}_\theta(r_H)\equiv4\,\pi \, G_d\,  M  2^{3-d}\left(\frac{r_H^2}{\theta}\right)^{\frac{d-3}{2}} \Bigg\{
 2^{\frac{d-3}{2}}r_H^4\left(\frac{r_H^2}{\theta}\right)^{-\frac{d+1}{4}}{e^{-{\frac {
 r_H^{2}}{ 8\theta}}}}{\cal{W}}\left(
\frac{d+1}{4},\frac{d+3}{4},{\frac {
  r_H ^{2}}{4\theta}} \right) $$$$\left[\left( \frac{d-1}{2} \right)
{\theta^2} \left( 1+d \right)  \left( d+3 \right)\right]^{-1}+
 \left( {\frac { r_H  ^{2}}{2\theta}}+1+d \right)
 \left( {\frac {  r_H  ^{2}}{\theta}} \right) ^{-
\frac{d+1}{4}}2^{\frac{d+1}{2}}{e^{-\,{\frac {  r_H
^{2}}{8\theta}}}}$$$${\cal{W}}\left(
\frac{d+5}{4},\frac{d+3}{4},{\frac {  r_H  ^ {2}}{4\theta}} \right)
\left[\left( \frac{d-1}{2} \right)  \left( 1+d \right) \right]^{-1}
\Bigg\} \left( \left( d-2 \right) {\pi }^{\frac{d-1}{2}}
\right)^{-1}.$$ Where ${\cal{W}}(\mu,\nu,z)$ shows the {\it
Whittaker function} defined in terms of the {\it hypergeometric
functions} as follows
$${\cal{W}}(\mu,\nu,z)=hypergeom\Big(\nu-\mu+\frac{1}{2}\, ,\,
2\nu+1\, ,\, z\Big)\,e^{-\frac{z}{2}}z^{\nu+\frac{1}{2}}.$$
Analytical solution of the Eq.~(\ref{mat:8}) for $r_{H}$ in a closed
form is impossible, but we can approximately solve it by setting
$r_H=2M$ into the lower incomplete Gamma function of the
Eq.~(\ref{mat:4}) to find {\footnote{We will use this approximation
in the next section too.}}
$$r_H=\Big(2{\cal{M}}_\theta(r_H)\Big)^{\frac{1}{d-3}}\approx
\Bigg\{\frac{8\,\pi \, G_d\,  M ^{d}}{\theta^{\frac{d-3}{2}}} \Bigg[
4\, \frac{ M ^{-\frac{d-3}{2}}}{\theta^{-\frac{d+1}{4}}}{e^{-{\frac
{  M ^{2}}{ 2\theta}}}}{\cal{W}}\left(
\frac{d+1}{4},\frac{d+3}{4},{\frac {
  M ^{2}}{\theta}} \right) $$$$\left(\left( \frac{d-1}{2} \right)
{\theta} \left( 1+d \right)  \left( d+3 \right)\right)^{-1}+
 \left( 2\,{\frac { M  ^{2}}{\theta}}+1+d \right)
\theta\, \left( {\frac {  M  ^{2}}{\theta}} \right) ^{-
\frac{d+1}{4}}{e^{-\,{\frac {  M
^{2}}{2\theta}}}}$$\begin{equation}{\cal{W}}\left(
\frac{d+5}{4},\frac{d+3}{4},{\frac {  M  ^ {2}}{\theta}} \right)
\left(\left( \frac{d-1}{2} \right)  M  ^{2} \left( 1+d \right)
\right)^{-1} \Bigg] \left( \left( d-2 \right) {\pi }^{\frac{d-1}{2}}
{\theta}\right)^{-1}\Bigg\}^{\frac{1}{d-3}}.\label{mat:9}
\end{equation}
For very large masses, one can easily recover the classical
Schwarzschild radius in a spacetime with $d$ dimensions,
$r_H\approx\left(2m\right)^{\frac{1}{d-3}}$.

Now we calculate Hawking temperature of noncommutative black hole by
investigating its radiating behavior. By definition,
\begin{equation}
T_H={1\over {4\pi}} {{dg_{00}}\over {dr}}|_{r=r_H}.\label{mat:10}
\end{equation}
Black hole temperature for some arbitrary number of spacetime
dimensions can be calculated as follows
\begin{equation}T_4=G_4M\Bigg[\frac{{\cal{E}} \left( {\frac {r_H}{2\sqrt{\theta}}}
\right)}{2\pi
r_H^2}-\frac{r_He^{-\frac{r_H^2}{4\theta}}}{4(\pi\theta)^{\frac{3}{2}}}\bigg(1+\frac{2\theta}{r_H^2}\bigg)\Bigg],\label{mat:11}
\end{equation}
\begin{equation}T_5=G_5M\Bigg[\frac{4}{3\pi^2
r_H^3}-\frac{r_He^{-\frac{r_H^2}{4\theta}}}{12(\pi\theta)^2}\bigg(1+\frac{4\theta}{r_H^2}
+\frac{16\theta^2}{r_H^4}\bigg)\Bigg],\label{mat:12}
\end{equation}
\begin{equation}T_6=G_6M\Bigg[\frac{9{\cal{E}} \left( {\frac {r_H}{2\sqrt{\theta}}}
\right)}{8\pi^2
r_H^4}-\frac{r_He^{-\frac{r_H^2}{4\theta}}}{32(\pi\theta)^{\frac{5}{2}}}\bigg(1+\frac{6\theta}{r_H^2}
+\frac{36\theta^2}{r_H^4}\bigg)\Bigg],\label{mat:13}
\end{equation}
\begin{equation}T_7=G_7M\Bigg[\frac{32}{10\pi^3
r_H^5}-\frac{r_He^{-\frac{r_H^2}{4\theta}}}{80(\pi\theta)^3}\bigg(1+\frac{8\theta}{r_H^2}
+\frac{64\theta^2}{r_H^4}+\frac{256\theta^3}{r_H^6}\bigg)\Bigg],\label{mat:14}
\end{equation}
and so on.  ${\cal{E}}(x)$ shows the {\it Gauss Error Function}
defined as follows
$$ {\cal{E}}(x)\equiv \frac{2}{\sqrt{\pi}}\int_{0}^{x}e^{-t^2}dt.$$
While for all even values of $d$ Hawking temperature can be
expressed in terms of the Gauss error function, for odd values of
$d$ Hawking temperature has a closed analytical form. For very large
distances (or commutative case), the function
${\cal{E}}(\frac{r_H}{2\sqrt{\theta}})$ for even values of $d$ tends
to unity and other terms on the right hand side will be vanishing
exponentially. Therefore, one recovers the classical Hawking
temperature in $d$ dimensions
\begin{equation}T_H=2\left(\frac{d-3}{d-2}\right)\pi^{-\frac{d-1}{2}}G_d\,M\,
\Gamma\bigg(\frac{d-1}{2}\bigg)r_H^{2-d}.\label{mat:15}
\end{equation}
In this situation, if we want to obtain the simple $d$-dependent
form of the noncommutative Hawking temperature, then it can be
approximated by utilizing the Eq.~(\ref{mat:9}) as follows
\begin{equation}
T_H=\frac{d-3}{4\pi r_H}.\label{mat:16}
\end{equation}
We will come back to this relation later in this paper.

It should be stressed that, if instead of distribution
(\ref{mat:2}), we adopt other kinds of probability distributions,
only the smeared mass distribution ${\cal{M}}_\theta$ will change
and all general properties will be the same as above. For example,
we consider a Lorentzian distribution of smeared matter as follows
\footnote{Generally, the noncommutativity parameter $\theta'$
appeared in Eq.~(\ref{mat:17}) is not exactly the same as $\theta$
appeared in Eq.~(\ref{mat:2}).}
\begin{equation}
\rho_{\theta'}(r)=\frac{M\sqrt{\theta'}}{\pi^2(r^2+\theta')^{\frac{d}{2}}}.\label{mat:17}
\end{equation}
The smeared mass distribution is now given by
\begin{equation}
{\cal{M}}_{\theta'}=\frac{2m(d-2)!!}{\pi(d-3)!!}\int_0^{\frac{r}{\sqrt{\theta'}}}
\frac{t^{d-2}} {(1+t^2)^{\frac{d}{2}}}dt.\label{mat:18}
\end{equation}
In the limit of $\theta'\rightarrow0$, we obtain
${\cal{M}}_{\theta'}\rightarrow m$. The Lorentzian smeared mass,
${\cal{M}}_{\theta'}$, has limiting properties similar to Gaussian
one and therefore can be compared with Gaussian smeared mass
distribution, ${\cal{M}}_{\theta}$. Thus, most of the outcomes that
we obtained for Gaussian profile, stay applicable if we pick out a
different form of the probability distribution of smeared matter
{\footnote{The difference between various forms of smeared mass
distributions is considerable around $M < M_0$ where there are
sensitivity to noncommutative effects and detailed form of the
matter distribution. Therefore, we focus on situations that the
condition $M \geq M_0$ holds.}} (see also \cite{noz1,riz}).

As an important remark, we note that some authors have studied black
hole thermodynamics in the NC framework adopting a coordinate
noncommutativity against coherent state approach (see \cite{noz4}
and references therein). A question then arises: what is the
difference between these two approaches? The standard way to handle
NC problems is based on Moyal $\star$-product. That means to use
complex number commuting coordinates and shift noncommutativity in
the product between functions. This is mathematically valid, but it
is physically useless since any model written in terms of
$\star$-product, even the simplest field theory, is nonlocal and it
is not obvious how to handle nonlocal quantum field theory. One
suggested approach is perturbation in the $\theta$ parameter
\cite{chai}. This is physically reasonable due to the fact that once
expanded up to a given order in $\theta$, the resulting field theory
becomes local. The smeared picture of particles based on coordinate
coherent states defines complex number coordinates as quantum mean
values of the original noncommuting ones between coordinate coherent
states \cite{sma}. In other words, in this setup one can see
commuting coordinates as classical limit (in the quantum mechanical
sense) of the noncommuting ones. In this framework, the emergent
semiclassical geometry keeps memory of its origin. For example, free
propagation of a point-like object is described by a minimal width
Gaussian wave-packet as has been considered in our setup. So, the
difference between two approaches lies in the definition of quantum
field theoretical propagators.

\section{\label{sec:3}Particles' Tunneling Near the Smeared Quantum Horizon}
Now we provide a detailed calculation of quantum tunneling from the
NC black hole event horizon. To portray the NC quantum tunneling
procedure, where a particle moves in dynamical geometry and passes
through the horizon without singularity on the path, we should take
advantage of a coordinates system that, unlike Schwarzschild
coordinates, is not singular at the horizon. A convenient choice in
this regard is Painlev\'{e} coordinate transformation \cite{pai}
which is obtained by definition of a new NC time coordinate
\begin{equation}
dt=dt_s+\frac{\sqrt{2{\cal{M}}_\theta
r^{d-3}}}{r^{d-3}-2{\cal{M}}_\theta}dr,\label{mat:19}
\end{equation}
where $t_s$ is the Schwarzschild time coordinate. Note that only
Schwarzschild time coordinate is transformed, while radial
coordinate and angular coordinates remain unchanged. Now the NC
Painlev\'{e}-Schwarzschild metric in spacetime with $d$ dimensions
takes the following form
\begin{equation}
ds^2 =-\bigg(1 - \frac{2{\cal{M}}_\theta}{r^{d-3}} \bigg) dt^2 +
2\sqrt{\frac{2{\cal{M}}_\theta}{r^{d-3}}} dtdr + dr^2 + r^2
d\Omega_{(d-2)}^2.\label{mat:20}
\end{equation}
The metric in these new coordinates is stationary and non-static. In
addition, neither coordinate nor intrinsic singularities are
present. Radial null geodesics (the equation of motion for massless
particles) in this geometry are as follows
\begin{equation}
\dot{r}\equiv\frac{dr}{dt}=\pm
1-\sqrt{\frac{2{\cal{M}}_\theta}{r^{d-3}}},\label{mat:21}
\end{equation}
where the upper (lower) sign corresponds to an outgoing (ingoing)
null geodesic respectively. If we assume $t$ increases in the
direction of future, then the above equations will be corrected by
the particle's self-gravitation effect. Kraus and Wilczek \cite{kra}
scrutinized the motion of particles in the $s$-wave as spherical
massless shells in dynamical geometry and expanded self-gravitating
shells in Hamiltonian gravity (see also \cite{sha}). Supplementary
elaborations was accomplished by Parikh and Wilczek \cite{par1}. In
this manner, we are going to develop their framework to the NC
coordinate coherent states in a model universe with
extra-dimensions.

We set the total ADM mass ($M$) of the spacetime to be fixed, and
permit the hole mass to fluctuate. Also, we consider the reaction of
the background geometry to an emitted quantum of energy $E$ which
moves on the geodesics of a spacetime with $M$ replaced by $M-E$. In
this manner, we should replace $M$ by $M-E$ in Eqs.~(\ref{mat:19},
\ref{mat:20}, and \ref{mat:21}). The geodesic equation of motion,
Eq.~(\ref{mat:21}), can be written as
\begin{equation}
\dot{r}=\pm
1-\sqrt{\frac{2{\cal{M}}_\theta\left(M-E\right)}{r^{d-3}}}.\label{mat:22}
\end{equation}
Since the characteristic wavelength of the radiation near the
horizon is always small due to infinite blue-shift, the wave-number
approaches infinity. Hence the WKB approximation is valid near the
horizon. In the WKB limit, the tunneling probability or emission
rate for the classically forbidden trajectories as a function of
imaginary part of the particle action at stationary phase takes the
following form
\begin{equation}
\Gamma\sim\exp(-2\textmd{Im}\, I).\label{mat:23}
\end{equation}
To compute the imaginary part of the action, we consider a spherical
shell of massless particles that move on radial null geodesics.
Since the radial null geodesics can be interpreted as $s$-waves
outgoing positive energy particles which cross the horizon outward
from $r_{in}$ to $r_{out}$, we use these radial null geodesics to
calculate the $\textmd{Im}\, I$, as follows \footnote{ Note that we
require $r_{in}>r_{out}$ where $r_{in}=\frac{16\pi G_d M}{(d-2)
\Omega_{(d-2)}}\frac{1}{\Gamma(\frac{d-1}{2})}\,\int_0^{\frac{r_{in}^2}{4\theta}}dt\,e^{-t}
\,t^{(\frac{d-3}{2})}$ \quad and \quad $r_{out}=\frac{16\pi G_d
(M-E)}{(d-2)
\Omega_{(d-2)}}\frac{1}{\Gamma(\frac{d-1}{2})}\,\int_0^{\frac{r_{out}^2}{4\theta}}dt\,e^{-t}
\,t^{(\frac{d-3}{2})}$.}
\begin{equation}
\textmd{Im}\,
I=\textmd{Im}\int_{r_{in}}^{r_{out}}p_rdr=\textmd{Im}\int_{r_{in}}^{r_{out}}\int_0^{p_r}dp'_rdr.\label{mat:24}
\end{equation}
Therefore, we incorporate back-reaction effects in a finite
separation between initial and final radius as a result of
self-gravitation effects of outgoing shells. On the other hand,
according to energy conservation, the tunneling barrier is produced
by a change in the radius (reduction of black hole horizon) just by
emitting particles. We now change the integration variable from
momentum to energy by applying Hamilton's equation of motion,
$\dot{r}=\frac{dH}{dp_r}|_r$ . So, we find
\begin{equation}\textmd{Im}\,
I=\textmd{Im}\int_{M}^{M-E}\int_{r_{in}}^{r_{out}}\frac{dr}{\dot{r}}dH,\label{mat:25}
\end{equation}
where the Hamiltonian is given by $H=M-E'$. The $r$-integral can be
solved firstly by deforming the contour so that
\begin{equation}\textmd{Im}\,
I=\textmd{Im}\int_{0}^{E}\int_{r_{in}}^{r_{out}}\frac{dr}
{1-\sqrt{\frac{2\,{\cal{M}}_\theta\left(M-E'\right)}{r^{d-3}}}}(-dE').\label{mat:26}
\end{equation}
The $r$-integral has a pole at the horizon in which it lies on the
line of integration and leads to ($-\pi i$) times the residue
\begin{equation}\textmd{Im}\,I=\textmd{Im}\int_{0}^{E}(-\pi
i)\left(\frac{2}{d-3}\right)\Big(2{\cal{M}}_\theta\left(M-E'\right)\Big)^{\frac{1}{d-3}}(-dE').\label{mat:27}
\end{equation}
The above integration can be preformed without writing out the
explicit form of radial null geodesics. In the vicinity of the
horizon, $\dot{r}$ behaves as
\begin{equation}
\dot{r}\simeq(r-r_H)\kappa(M)+O\left((r-r_H)^2\right),\label{mat:28}
\end{equation}
where $\kappa(M)$ is the horizon surface gravity. The mentioned
integration can be done simply by inserting the relation
(\ref{mat:28}), in which self-gravitation is included, into the
integral (\ref{mat:25}) and performing the $r$-integral as follows
\begin{eqnarray}\textmd{Im}\,
I=-\textmd{Im}\int_{0}^{E}\int_{r_{in}}^{r_{out}}\frac{drdE'}{(r-r_H)\kappa(M-E')}
=\pi\int_{0}^{E}\frac{dE'}{\kappa(M-E')}.\label{mat:29}
\end{eqnarray}
Comparing two integrands in the relations (\ref{mat:27}) and
(\ref{mat:29}) gives us
\begin{equation}
\kappa(M-E)=\frac{d-3}{2\left(2{\cal{M}}_\theta(M-E)\right)^{\frac{1}{d-3}}}.\label{mat:30}
\end{equation}
Now, Hawking temperature is given by
\begin{equation}
T_H=\frac{\kappa(M)}{2\pi}\approx\frac{d-3}{4\pi\left(2{\cal{M}}_\theta(M)\right)^{\frac{1}{d-3}}}.\label{mat:31}
\end{equation}
It is clear that, two Eqs.~(\ref{mat:16}) and (\ref{mat:31}) for the
NC black hole temperature coincide, but note that the approach to
achieve (\ref{mat:31}) uses some approximations. This coincidence
confirms that our calculations are approximately correct.

Here, we would like to obtain the radial geodesics of the massive
particles which are different with massless case. Recently, some
authors have extended Parikh-Wilczek work to the massive particles'
tunneling ~\cite{zha1,zha2,jia}. In this regard, once again we
follow the line element of the NC Painlev\'{e}-Schwarzschild black
hole, that is
\begin{equation}
ds^2 =g_{00}dt^2+2g_{01}dtdr+g_{11}dr^2+ r^2
d\Omega_{(d-2)}^2.\label{mat:m1}
\end{equation}
This line-element can be obtained by definition of a noncommutative
time coordinate transformation as
\begin{equation}
dt=dt_s+dt_{syn},\label{mat:m2}
\end{equation}
where
\begin{equation}
dt_{syn}=\frac{\sqrt{2{\cal{M}}_\theta
r^{d-3}}}{r^{d-3}-2{\cal{M}}_\theta}dr=-\frac{g_{01}}{g_{00}}dr.\label{mat:m3}
\end{equation}
This relation, which has been written based on the Landau's theory
of the synchronization of clocks \cite{lan}, allows us to
synchronize clocks in any infinitesimal radial positions of
particles ($d\Omega_{(d-2)}=0$). Since the tunneling phenomena
through the quantum horizon, the barrier, is an instantaneous
procedure, it is important to apply Landau's theory of the
coordinate clock synchronization to the tunneling process. The
mechanism for tunneling through the quantum horizon is that particle
anti-particle pair is created at the event horizon. So, we have two
events that occur simultaneously; one event is anti-particle and
tunnels into the barrier but the other particle tunnels out the
barrier. In fact, the relation (\ref{mat:m3}) mentions the
difference of coordinate times for these two simultaneous events
occurring at infinitesimally adjacent radial positions. According to
the non-relativistic quantum theory, de Broglie's hypothesis and the
WKB approximation, it can be easily shown that the treatment of the
massive particle's tunneling as a massive shell is approximately
derived by the phase velocity $v_p$ of the de Broglie s-wave. The
relationship between phase velocity $v_p$ and group velocity $v_g$
is given by \cite{zha1,zha2,jia}
\begin{equation}
v_p=\dot{r}=\frac{1}{2}v_g.\label{mat:m4}
\end{equation}
In the case of $d\Omega_{(d-2)}=0$, according to the relation
(\ref{mat:m3}), the group velocity is
\begin{equation}
v_g=-\frac{g_{00}}{g_{01}}.\label{mat:m5}
\end{equation}
Thus, the outgoing motion of the massive particles take the
following form
\begin{equation}
\dot{r}=-\frac{g_{00}}{2g_{01}}=
\frac{r^{d-3}-2{\cal{M}}_\theta}{2\sqrt{2{\cal{M}}_\theta
r^{d-3}}}.\label{mat:m6}
\end{equation}
We should incorporate back-reaction effects in this relation and
then the result should be inserted into relation (\ref{mat:25}) to
obtain a relation corresponding to (\ref{mat:27}) but now for
massive particles. Let us come back to the integral (\ref{mat:27}).
This integral has not a closed analytical solution, so we are going
to compute Taylor series expansion of integrand, with respect to the
particle's energy $E$, about the zero point and just to first order.
Then we can solve the integral to find
$$\textmd{Im}\,I= \pi \, \Big[ 16\,\pi \, G_d\,M{\pi }^{-\frac{d-1}{2}} \Big( {
\frac {{M}^{2}}{\theta}} \Big) ^{\frac{d-1}{2}} e^{- \frac
{M^2}{\theta}} \Big( 4\,{M}^{4}{\cal{H}} +
2\,d\theta\,{M}^{2}+6\,\theta\,{M}^{2}+4\,d{\theta}^{2}+3\,{\theta}^{2
}+{d}^{2}{\theta}^{2} \Big)  ( d-2 ) ^{-1}$$$${\theta}^{-2}
 ( d-1 ) ^{-1} ( 1+d ) ^{-1} ( d+3 )
^{-1} \Big] ^{ \frac{1}{d-3}}E \Big( -E{d}^{4}{\theta}^{
3}-180\,{M}^{3}{\theta}^{2}-90\,M{\theta}^{3}-4\,E{M}^{4}{\cal{H}}\theta\,{d}^{2}-16\,E{M}^{6}{\cal{H'}}
+$$$$40\,E{M}^{6}{\cal{H}} +2\,M{\theta}^{3}{d}
^{4}+4\,{M}^{3}{\theta}^{2}{d}^{3}-120\,{M}^{5}\theta\,{\cal{H}}
+20\,{M}^{3
}{\theta}^{2}{d}^{2}-36\,{M}^{3}{\theta}^{2}d+12\,M{\theta}^{3}{d}^{3}
-8\,M{\theta}^{3}{d}^{2}+$$$$16\,{M}^{5}\theta\,d{\cal{H}}
+8\,{M}^{5}\theta\,{d }^{2}{\cal{H}} +8\,E{M}^{6}{\cal{H}}
d-80\,E{M}^{4}{\cal{H}} \theta-16\,Ed{\theta}^{2}{M
}^{2}-2\,E{d}^{2}{\theta}^{2}{M}^{2}+32\,E\theta\,d{M}^{4}+$$$$4\,E\theta
\,{d}^{2}{M}^{4}-108\,M{\theta}^{3}d+60\,E\theta\,{M}^{4}-30\,E{\theta
}^{2}{M}^{2}-15\,Ed{\theta}^{3}-23\,E{d}^{2}{\theta}^{3}-9\,E{d}^{3}{
\theta}^{3}-36\,E{M}^{4}{\cal{H}} \theta\,d \Big)$$
 \begin{equation}  ( d-3)
^{-2}{\theta}^{-1} ( 5+d ) ^{-1}{M}^{-1} \Big( 4 \,{M}^{4}{\cal{H}}
+2\,d\theta\,{M}^{2}+6\,\theta\,{M}^{2}+4\,d{\theta}^
{2}+3\,{\theta}^{2}+{d}^{2}{\theta}^{2} \Big) ^{-1},\label{mat:32}
\end{equation}
where ${\cal{H}}={\it hypergeom}\left( 1 ,\frac{5+d}{2},{\frac
{{M}^{2}}{\theta}} \right)$\, and \,${\cal{H'}}={\it hypergeom}
\left( 2 ,\frac{7+d}{2},{ \frac {{M}^{2}}{\theta}} \right)$. Since,
both particle and anti-particle (corresponding to a time reversed
state that can be seen backward in time by replacing
$\sqrt{\frac{2\cal{M}_\theta}{r^{d-3}}}$ by
$-\sqrt{\frac{2\cal{M}_\theta}{r^{d-3}}}$ in the metric
(\ref{mat:20})) anticipate in the emission rate of the Hawking
process via tunneling with the same amounts, we have to add their
amplitudes firstly and then squaring it to obtain the emission
probability (Eq.~(\ref{mat:23})).

On the other side, using the first law of black hole thermodynamics,
$dM=\frac{\kappa}{2\pi}dS$, and Eq.~(\ref{mat:29}),\, one can find
the imaginary part of the action as follows \cite{par1,par2,par3}
(see also \cite{kes,med,majhi1,majhi2})
\begin{equation}
\textmd{Im}\,I=-\frac{1}{2}\int_{S_{NC}(M)}^{S_{NC}(M-E)}dS=-\frac{1}{2}\Delta
S_{NC},\label{mat:33}
\end{equation}
which shows that the tunneling approach concurs with the thermal
Hawking spectrum just in the limit of low energy. Hawking radiation
of black holes as quantum tunneling was also studied in the context
of string theory \cite{kes}, and it was demonstrated that the
tunneling probability in the high energy limit is analogous to a
difference between counting of states in the microcanonical and
canonical ensembles. It can be easily seen that our computed
tunneling probability to first order in $E$ generates the Boltzmann
factor in the canonical ensemble $\Gamma\sim\exp(-\beta E)$, which
is characterized by the inverse temperature as the coefficient
$\beta$. So, the tunneling probabilities in the high energy limit
are proportional to $\exp(\Delta S)$ (see also \cite{mas}),
\begin{eqnarray}
\Gamma\sim\exp(-2\textmd{Im}\,
I)\sim\frac{e^{S_{final}}}{e^{S_{initial}}}=\exp(\Delta
S)=\exp[S(M-E)-S(M)],\label{mat:34}
\end{eqnarray}
where $\Delta S$ is the difference in black hole entropies before
and after emission. This means that at higher energies the emission
rate depends on the final and initial number of accessible
microstates of the system. Therefore at higher energies the emission
spectrum cannot be exactly thermal since the high energy
modifications flow from the physics of energy conservation with
noncommutativity corrections. In fact, to linear order in $E$, two
expressions for $\Gamma$ in the microcanonical and canonical
ensembles coincide. Thus, obviously the emission rate (\ref{mat:34})
deviates from the pure thermal emission but is corresponding to an
underlying unitary quantum theory \cite{par2,par3}.

At this stage, one can easily show that there are no correlations
between emitted particles even with incorporation of the
noncommutativity modifications in an extra dimensional setup. If
there is any correlation between different emitted modes, at least
part of the information coming out of the black hole to be retained
in these correlations. As our analysis shows, at least up to second
order in particle's energy, there will be no correlation between the
different modes of radiation. This reflects the fact that
information does not come out continuously during the evaporation
process at least at late-times {\footnote{Our calculations are
performed based on Eq.~(\ref{mat:23}) in stationary phase, hence the
phrase {\it late-time}.}}. This indicates that the tunneling
probability of two particles of energy $E_1$ and $E_2$ is comparable
to the tunneling probability of one particle with composed energy,
$E=E_1+E_2$, {\it i.e.}
\begin{equation}
\Delta S_{E_1}+\Delta S_{E_2}=\Delta S_{(E_1+E_2)}\Rightarrow
\chi(E_1+E_2;E_1,E_2)=0.\label{mat:35}
\end{equation}
It can be numerically confirmed that these probabilities of emission
are actually uncorrelated. On the other hand, the statistical
correlation function $\chi(E;E_1,E_2)$, is zero which leads to the
independence between different modes of radiation during the
evaporation. Accordingly, space noncommutativity and higher
dimensional considerations are not enough to provide the basis for
existence of the correlations between different modes.

There are four main proposals about what happens to the information
that falls into a black hole. Based on the first proposal, the black
hole can evaporate completely as uncorrelated thermal radiation in
each mode and all the information including the original quantum
state that formed the black hole (excluding its mass, charge and
angular momentum), would disappear from our universe. However, this
proposal allows the pure states to evolve into the mixed states,
which is incompatible with the basic principles of quantum
mechanics. Second proposal is that the black hole can completely
disappear, but the information appears in the final burst of
radiation when the black hole shrinks to the Planck size. A third
possibility is that the information comes out in non-thermal
correlations within the Hawking radiation, the process being
portrayed by a unitary $S$-matrix. In other words, there are
non-thermal correlations between different modes of radiation
throughout the evaporation process that information emerges
ceaselessly encoded through them. And fourth idea is that the black
hole never disappears completely, and the information is not lost,
but would be stored in a Planckian-size stable remnant. In this
section, we have studied the credibility of the third and fourth
conjectures within a NC spacetime feature. We have shown that
inclusion of noncommutativity and braneworld effects are not enough
to create correlations between emitted modes of black hole
evaporation. However, due to spacetime noncommutativity, the
information might be preserved by a stable black hole remnant. In
our opinion, this is one of the successful scenarios for information
loss paradox currently \cite{noz1}. In next section we introduce a
procedure to create correlations between emitted modes.

\section{\label{sec:4}Quantum Gravity as a Source of Correlations}

All quantum gravity scenarios support the idea that there is a
fundamental length scale that cannot be probed. For instance, in
string theory there is a restraint in probing distances smaller than
the string length. Hence the standard uncertainty principle is
modified to include this restricted resolution of the spacetime
points. The result of this modification is known as the Generalized
Uncertainty Principle (GUP) which is actually a representation of
the quantum nature of spacetime at Planck scale. In a model universe
with extra dimensions, GUP can be generalized to the tensorial form
as follows
\begin{equation}
\Delta x_i\geq\frac{1}{\Delta p_i}+\alpha L_{Pl}^{2}\Delta
p_i,\label{mat:35-1}
\end{equation}
where the fundamental Planck length is defined as $L_{Pl}=
G_d^{\frac{1}{d-2}}$ and $\alpha$ is a dimensionless constant of the
order of unity that depends on the specific model. The above GUP can
be acquired in the context of string theory \cite{mini1}, NC quantum
theory \cite{mini2}, loop quantum gravity \cite{mini3}, or from
black hole gedanken experiments \cite{mini4}. In the standard limit,
$\Delta x_i\gg L_{Pl}$, it yields the ordinary uncertainty
principle, $\Delta x_i\Delta p_i\geq1$. The second term in {\it
r.h.s} of Eq.~(\ref{mat:35-1}) becomes considerable when the
momentum and distance scales approach the Planck scale. For a
spherical black hole the thermodynamic quantities can be obtained in
a heuristic manner using the standard uncertainty principle
\cite{sup}. Application of the GUP to black hole thermodynamics in
the same manner, modifies the results dramatically by incorporation
of quantum gravity effects in the final stages of black hole
evaporation with a rich phenomenology \cite{adl}. In this section we
use GUP to find an alternative to retrieve the lost information in
the black hole evaporation process. This alternative is related to
the correlations between different modes of evaporation. Can GUP
provide the required correlations? This is an important question
which we answer in a tunneling framework. A first effort in this
direction was given by Arzano {\it et al} \cite{med}, resulting to
the quantum corrected entropy with an additional logarithmic
correction term. However, such a formulation does not prepare a way
to find a non-zero statistical correlation function. In this section
we follow a different way to retrieve the information. In a recent
paper \cite{hamid}, we investigated the modifications of the Hawking
radiation by the tunneling process and the GUP. By using the
GUP-corrected de Broglie wave length, the squeezing of the
fundamental momentum-space cell, and consequently a GUP-corrected
energy, we find the non-thermal effects which lead to a non-zero
statistical correlation function between probabilities of tunneling
of two particles with different energies. In this section, we extend
our framework to model universe with extra dimensions.

Amelino-Camelia {\it et al}  \cite{amelino} (see also \cite{sefid})
studied the black hole evaporation process after an analysis of the
GUP-induced modification of the black body radiation spectrum. If
GUP is fundamental concept of quantum gravity, it should appear in
de Broglie relation as follows
\begin{equation}
\lambda\simeq\frac{1}{p}\left(1+\alpha
L_{Pl}^2p^2\right),\label{mat:36}
\end{equation}
or
\begin{equation}
{\cal{E}}\simeq E(1+\alpha L_{Pl}^2E^2).\label{mat:37}
\end{equation}
There are other compelling reasons from NC geometry and loop quantum
gravity that support relation (\ref{mat:37}) (see for instance
\cite{amelino,sefid} and references therein). Now, we consider a
massless particle {\it i.e.} a shell and take into consideration the
response of the background geometry to a radiated quantum of energy
$E$ with GUP correction {\it i.e.} ${\cal{E}}$. The particle moves
on the geodesics of a spacetime with $M-{\cal{E}}$ substituted for
$M$. With these preliminaries and also considering the deformed
Hamilton's equation of motion as
\begin{equation}
\dot{r}\simeq\left(1+\alpha L_{Pl}^2
{\cal{E}}^2\right)\frac{dH}{dp_r}|_r,\label{mat:38}
\end{equation}
eventually the imaginary part of the action takes the following form
\begin{equation}
\textmd{Im}\,I=\textmd{Im}\int_{0}^{\cal{E}}(-\pi
i)\left(\frac{2}{d-3}\right)\left(1+\alpha L_{Pl}^2
{\cal{E'}}^2\right)\Big(2\left(M-{\cal{E'}}\right)\Big)^{\frac{1}{d-3}}(-d{\cal{E'}}).\label{mat:39}
\end{equation}
The leading order correction is proportional to the square of
$(\sqrt{\alpha}~L_p)$. This integral, similar to the integral
(\ref{mat:27}), has not a closed analytical solution, so we are
going to compute Taylor series expansion of integrand (with respect
to the particle's energy $E$) about the zero point and just to first
order. Then we can solve the integral to find
$$\textmd{Im}\,I=\frac{\pi}{(d-3)^2}\Bigg[(d-3)\left(2^{d-2}M\right)^{\frac{1}{d-3}}E-\left(2M^{4-d}
\right)^{\frac{1}{d-3}}E^2+$$\begin{equation}+\alpha
L_{Pl}^2E^3\left((d-3)\left(2^{d-2}M\right)^{\frac{1}{d-3}}-2\left(2M^{4-d}
\right)^{\frac{1}{d-3}}E\right)+O(\alpha^2L_{Pl}^4)\Bigg].\label{mat:40}
\end{equation}
The tunneling rate is therefore
$$\Gamma\sim\exp\Bigg(-\frac{2\pi}{(d-3)^2}\Bigg[(d-3)\left(2^{d-2}M\right)^{\frac{1}{d-3}}E-\left(2M^{4-d}
\right)^{\frac{1}{d-3}}E^2+$$\begin{equation}+\alpha
L_{Pl}^2E^3\left((d-3)\left(2^{d-2}M\right)^{\frac{1}{d-3}}-2\left(2M^{4-d}
\right)^{\frac{1}{d-3}}E\right)+O(\alpha^2L_{Pl}^4)\Bigg]\Bigg).\label{mat:41}
\end{equation}
For $d = 4$, the first and second expressions in the exponential
exhibit non-thermal behavior that was discovered firstly in
Ref.~\cite{par1}. So, the emission spectrum cannot be strictly
thermal. In our situation, there is an additional term depending on
the GUP parameter in first order and cannot be
neglected once the black hole mass becomes comparable to the Planck mass.\\
We now illustrate whether or not, the emission rates for the
different modes of radiation during the evaporation are mutually
related from a statistical viewpoint. Using (\ref{mat:41}), the
emission rate for a first quantum with energy $E_1$, gives
$$\ln\Gamma_{E_1}=-\frac{2\pi}{(d-3)^2}\Bigg[(d-3)\left(2^{d-2}M\right)^{\frac{1}{d-3}}E_1-\left(2M^{4-d}
\right)^{\frac{1}{d-3}}E_1^2+$$\begin{equation}+\alpha
L_{Pl}^2E_1^3\left((d-3)\left(2^{d-2}M\right)^{\frac{1}{d-3}}-2\left(2M^{4-d}
\right)^{\frac{1}{d-3}}E_1\right)\Bigg].\label{mat:42}
\end{equation}
Similarly, the emission rate for a second quantum $E_2$, takes the
form
$$\ln\Gamma_{E_2}=-\frac{2\pi}{(d-3)^2}\Bigg[(d-3)\left(2^{d-2}(M-E_1)\right)^{\frac{1}{d-3}}E_2-\left(2(M-E_1)^{4-d}
\right)^{\frac{1}{d-3}}E_2^2+$$\begin{equation}+\alpha
L_{Pl}^2E_2^3\left((d-3)\left(2^{d-2}(M-E_1)\right)^{\frac{1}{d-3}}-2\left(2(M-E_1)^{4-d}
\right)^{\frac{1}{d-3}}E_2\right)\Bigg].\label{mat:43}
\end{equation}
Correspondingly, the emission rate for a single quantum with the
same total energy, $E=E_1+E_2$, yields
$$\ln\Gamma_{(E_1+E_2)}=-\frac{2\pi}{(d-3)^2}\Bigg[(d-3)\left(2^{d-2}M\right)^{\frac{1}{d-3}}(E_1+E_2)-\left(2M^{4-d}
\right)^{\frac{1}{d-3}}(E_1+E_2)^2+$$\begin{equation}+\alpha
L_{Pl}^2(E_1+E_2)^3\left((d-3)\left(2^{d-2}M\right)^{\frac{1}{d-3}}-2\left(2M^{4-d}
\right)^{\frac{1}{d-3}}(E_1+E_2)\right)\Bigg].\label{mat:44}
\end{equation}
It can be seen easily that these probabilities are correlated. On
the other hand, the non-zero statistical correlation function is
$$
\chi(E_1+E_2;E_1,E_2)=\frac{2\pi
E_2}{(d-3)^2}\Bigg[(d-3)\left[\left(2^{d-2}(M-E_1)\right)^{\frac{1}{d-3}}-\left(2^{d-2}M\right)^{\frac{1}{d-3}}\right]+
$$$$+E_2\left[\left(2M^{4-d}\right)^{\frac{1}{d-3}}-\left(2(M-E_1)^{4-d}\right)^{\frac{1}{d-3}}\right]+
E_1\left(2^{d-2}M^{4-d}\right)^{\frac{1}{d-3}}+$$$$+\alpha
L_{Pl}^2\Bigg(4E_1\left(2M^{4-d}\right)^{\frac{1}{d-3}}\left[2E_1^2+2E_2^2+3E_1E_2\right]-(d-3)\left(2^{d-2}M\right)^{\frac{1}{d-3}}
\left[3E_1^2+E_2^2+3E_1E_2\right]+$$\begin{equation}+(d-3)E_2^2\left(2^{d-2}(M-E_1)\right)^{\frac{1}{d-3}}+
E_2^3\left[\left(2^{d-2}M^{4-d}\right)^{\frac{1}{d-3}}-\left(2^{d-2}(M-E_1)^{4-d}\right)^{\frac{1}{d-3}}\right]\Bigg)\Bigg].\label{mat:45}
\end{equation}
Note that three first terms in square bracket in r.h.s. can be
neglected for any reasonable values of quantities. Appearance of
these terms originates from approximations used in the calculation
of integral in equation (\ref{mat:39}). Therefore we find
$$
\chi(E_1+E_2;E_1,E_2)=\frac{2\pi\alpha L_{Pl}^2
E_2}{(d-3)^2}\Bigg(4E_1\left(2M^{4-d}\right)^{\frac{1}{d-3}}\left[2E_1^2+2E_2^2+3E_1E_2\right]-$$$$-(d-3)\left(2^{d-2}M\right)^{\frac{1}{d-3}}
\left[3E_1^2+E_2^2+3E_1E_2\right]+(d-3)E_2^2\left(2^{d-2}(M-E_1)\right)^{\frac{1}{d-3}}+
$$\begin{equation}+E_2^3\left[\left(2^{d-2}M^{4-d}\right)^{\frac{1}{d-3}}-\left(2^{d-2}(M-E_1)^{4-d}\right)^{\frac{1}{d-3}}\right]\Bigg).\label{mat:46}
\end{equation}
This means that not only the probability of tunneling of two
particles of energy $E_1$ and $E_2$ is not the same as probability
of tunneling of one particle with their compound energies,
$E_1+E_2$, but also there are correlations between them. In fact,
whenever one quantum of emission is radiated from the surface of the
black hole horizon, the perturbations are created on the Planck
scale that influence the second quantum of emission, and these
perturbations cannot be neglected. Particularly, once the black hole
mass becomes comparable with the Planck mass, one cannot neglect the
effects of these perturbations. Therefore, as expected in
Ref.~\cite{med}, in this way the form of the amendments as
back-reaction effects by incorporation of GUP effects, are
sufficient by themselves to create correlation and therefore
recovery of information. So, at least part of the information leaks
out from the black hole as the non-thermal GUP correlations and this
provides a basis to solve information loss problem. The question
which arises here is time evolution of these correlations. This
problem is currently under investigation.

\section{\label{sec:5}Summary}
In this paper, generalization of the standard Hawking radiation via
quantum tunneling through the smeared quantum horizon of the
non-commutative Schwarzschild black hole has been studied. This
study is based on the solution of the Eq.~(\ref{mat:23}) in the
context of coordinate coherent state noncommutativity and in a model
universe with extra dimensions. The corrections of the emission
rates for both massless and massive particles have been performed.
We have shown that noncommutativity and extra-dimensional
considerations are not enough to create correlations between emitted
modes. That is, there are no correlations between the tunneling
rates of the different modes in the black hole radiation spectrum at
least at late-times and just to second order in $E$. However,
spacetime noncommutativity is adequate by itself to preserve
information due to the fact that in the NC framework, the black hole
does not evaporate completely and this leads to the existence of a
minimal non-zero mass (e.g., {\it a Planck-sized remnant}) in which
black hole can reduce to it at final stage of evaporation. So, part
of information might be preserved in this remnant. If one really
believes the idea of stable black hole remnants due to the fact that
there are some exact continuous global symmetries in the nature
\cite{bek}, then, based on our findings (Sections \ref{sec:2} and
\ref{sec:3}), we should accept this possibility that a part of
information stays inside the black hole and can be retained by a
stable Planck-sized remnant {\footnote{Of course, this assumption is
acceptable if information is really conserved in our world.}}. On
the other hand, the generalized uncertainty principle of string
theory or noncommutative quantum mechanics brings significant
modifications to the Hawking radiation. As it is well-known, the GUP
leads to a Planck-scale remnant at the endpoint of evaporation
process. We have shown that the GUP as modification of the de
Broglie relation in the quantum tunneling framework of the black
hole evaporation, leads to correlation between emitted modes of
evaporation. Hence in this setup, a part of the information leaks
out of the black hole in the form of non-thermal GUP correlations,
and the other part reside in the remnant. These features have the
potential to answer some questions regarding the black hole
information loss paradox and provide a more realistic background for
treating the black hole evaporation process, especially in its final
stages of evaporation. We stress that time-evolution of GUP-induced
correlations is an important issue which needs further investigation
and probably shed more light on information loss problem.\\

\acknowledgments{This work has been supported partially by Research
Institute for Astronomy and Astrophysics of Maragha, Iran.}

\appendix*

\section{}
The imaginary part of the action (Eq.~(\ref{mat:32})) with some
arbitrary number of dimensions can be calculated as follows
$$d=4\longrightarrow\textmd{Im}\, I=2\,\sqrt {\pi }
G_4\,{e^{-{\frac {{M}^{2}}{\theta}}}}E \Bigg(
-4\,E\frac{{M}^{3}}{\theta^{\frac{3}{2}}}-4
\,\frac{M^2}{\sqrt{\theta}}+2\,E \frac{M}{\sqrt{\theta}}+\sqrt {\pi
}{\cal{E}} \left( {\frac {M}{\sqrt{\theta}}} \right) {e^{{\frac {
{M}^{2}}{\theta}}}}\Big[2M-E\Big] \Bigg)
$$
$$d=5\longrightarrow\textmd{Im}\, I=\Bigg[\frac{\pi}{6}G_5M{e^{-{\frac
{{M}^{2}}{\theta}}}}\bigg(-\frac{M^2}{\theta}-1+{e^{{\frac
{{M}^{2}}{\theta}}}}\bigg)\Bigg]^{\frac{1}{2}}\bigg(M^3\theta{e^{-{\frac
{{M}^{2}}{\theta}}}}+M\theta^2{e^{-{\frac
{{M}^{2}}{\theta}}}}-M\theta^2\bigg)^{-1}$$$$E{e^{-{\frac
{{M}^{2}}{\theta}}}}\bigg(2EM^4+4M^3\theta-E\theta
M^2+4M\theta^2-4M\theta{e^{{\frac
{{M}^{2}}{\theta}}}}-E\theta^2+E\theta^2{e^{{\frac
{{M}^{2}}{\theta}}}}\bigg)
$$
$$d=6\longrightarrow\textmd{Im}\, I=\frac{\sqrt{\pi}}{18}2^{\frac{2}{3}}\Bigg(G_6Me^{-\frac{M^2}{\theta}}\bigg(
-4\frac{M^3}{\theta^{\frac{3}{2}}}-6\frac{M}{\sqrt{\theta}}+3\sqrt{\pi}{\cal{E}}
\left( {\frac {M}{\sqrt{\theta}}}
\right)e^{\frac{M^2}{\theta}}\bigg)\Bigg)^{\frac{1}{3}}$$$$\,E\Bigg(-8E\frac{M^5}{\sqrt{\theta}}
-24M^4\sqrt{\theta}+4EM^3\sqrt{\theta}-36M^2\theta^{\frac{3}{2}}+6EM\theta^{\frac{3}{2}}+\sqrt{\pi}\theta^2
{\cal{E}} \left( {\frac {M}{\sqrt{\theta}}}
\right)e^{\frac{M^2}{\theta}}\Big[18M-3E\Big]\Bigg)$$$$
\Bigg(-4M^4\sqrt{\theta}-6M^2\theta^{\frac{3}{2}}+3M\sqrt{\pi}\theta^2
{\cal{E}} \left( {\frac {M}{\sqrt{\theta}}}
\right)e^{\frac{M^2}{\theta}}\Bigg)^{-1}
$$
and so on.

\end{document}